\documentstyle[twocolumn,aps]{revtex}
\input epsf
\begin{document}
\draft
\twocolumn[\hsize\textwidth\columnwidth\hsize\csname@twocolumnfalse\endcsname

\title{Symmetry alteration of ensemble return distribution 
in crash and rally days of financial markets}
\author{Fabrizio Lillo and Rosario N. Mantegna}
\address{
Istituto Nazionale per la Fisica della Materia, Unit\`a di Palermo\\
and\\ Dipartimento di Fisica e Tecnologie Relative, Universit\`a 
di Palermo, Viale delle Scienze, I-90128, Palermo, Italia}
\maketitle
\begin{abstract}

We select the $n$ stocks traded in the New York Stock 
Exchange and we form a statistical ensemble of daily stock returns 
for each of the $k$ trading days of our database from the stock 
price time series. We study the ensemble return distribution for 
each trading day and we find that the symmetry properties of the 
ensemble return distribution drastically change in crash and rally 
days of the market. We compare these empirical results with numerical 
simulations based on the single-index model and we conclude that this 
model is unable to explain the behavior of the market in extreme days.

\end{abstract}
\vskip2pc]


In the last few years physicists interested in financial analysis 
have performed several empirical researches investigating the 
statistical properties of stock price and volatility time series 
of a single asset (or of an index) at fixed or at different time 
horizons \cite{Ms99,Palermo}. Other researches have been focusing
on the cross-correlation properties of simultaneously traded stocks
\cite{Mantegna99,Bouchaud99,Stanley99}. Another key aspect of the 
financial dynamics concerns the behavior
of the market in days of extreme gain or loss. Statistical properties 
at, before and immediately after extreme days have been recently 
investigated  by considering the behavior of market indices 
\cite{Sornette96,Chowdhury99}. In this letter we investigate 
extreme market days by following a different approach. Specifically,
 we investigate the 
return distribution of an ensemble 
of $n$ selected stocks simultaneously traded in a financial market in
market days of extreme crash or rally 
in the period of our database (from January 1987 to December 1998).

The investigation of the 
return distribution of an ensemble 
of stocks simultaneously traded was introduced in \cite{Lillo99}. 
The customary statistical properties of price return distribution of an 
ensemble of stocks are discussed elsewhere \cite{Lillo00}, here
we pose and answer the following question: 
Are crash and rally days significantly different from the 
typical market days with respect to the statistical properties of 
return distribution of an ensemble of stocks?


The investigated market is the New York Stock Exchange (NYSE) during the 
12-year period from January 1987 to December 1998 which corresponds to 3032
trading days. The total number of assets $n$ traded in NYSE is 
rapidly increasing and it ranges from $1128$ in 1987 to 
$2788$ in 1998. The total number of data records exceeds $6$ million. 

The variable investigated in our analysis is the daily price return, which is 
defined as
\begin{equation}
R_i(t)\equiv\frac{Y_i(t+1)-Y_i(t)}{Y_i(t)},
\end{equation}  
where $Y_i(t)$ is the closure price of $i-$th asset at day $t$ ($t=1,2,..$). 
In our study we 
consider only the trading days and we remove the weekends and the holidays 
from the data set. Moreover we do not consider price returns which are 
in absolute values greater than $50\%$ because some of these returns might 
be attributed to errors in the database and may affect in a considerable 
way the statistical analyses. We extract the $n$ returns of the 
$n$ stocks for each trading day and we consider the normalized 
probability density function (PDF) of price returns. 
The distribution of these returns gives an idea about the 
general activity of the market at the selected 
trading day. In the absence of extreme events, the central part of 
the distribution is conserved for long time periods.
In these periods the shape of the distribution is systematically 
non-Gaussian and approximately symmetrical \cite{Lillo00}.
We attribute the non-Gaussian profile of the central part of the
PDF to the presence of correlations among the stocks. Sometimes the 
PDF changes abruptly its shape either towards positive returns or 
towards negative returns. A systematic study of these days shows 
that they corresponds to extreme events in the market, i.e. to crash 
days and to rally days. In other words the periods in which the 
shape of the PDF changes corresponds to period of financial turmoil 
in the market. The most prominent example is the dramatic change of 
shape and of scale of the PDF observed during and after the 19 October 
1987 crash. Other dramatic changes are observed at the beginning 
of 1991 and at the end of 1998.
To illustrate in detail this behavior we consider the financial crisis
of October 1987. Figure 1 shows the surface and contour plot of the 
ensemble return PDFs determined in a $200$ trading days time interval
centered at 19th October 1987 (which correspond in abscissa to the 
arbitrary value 0). The $z$-axis is logarithmic in Figure 1.
The central part of the ensemble return distribution shows an
triangular-like shape which is approximately conserved far 
from the crisis.
\begin{figure}[t]
\epsfxsize=2.6in
\epsfbox{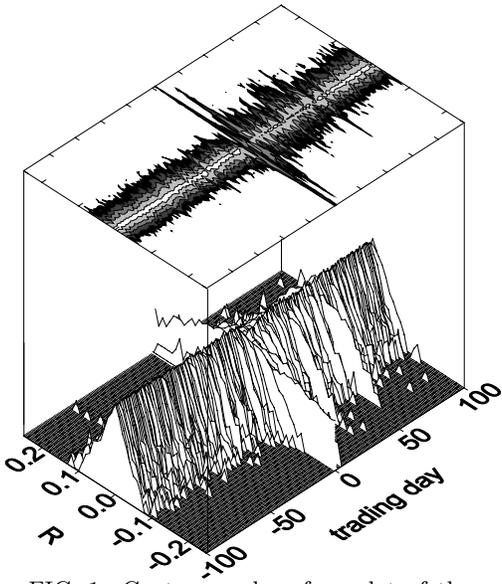}
\caption{Contour and surface plot of the ensemble return distribution in a 
200 trading days time interval centered at 19 October 1987 (corresponding to
0 in the abscissa). The probability density scale ($z$-axis) of the 
surface plot is logarithmic. The contour plot is obtained for equidistant
intervals of the logarithmic probability density. The brightest area 
of the contour plot corresponds to the 
most probable value.}
\label{fig1}
 \end{figure}
 At the crisis, the ensemble distribution moves 
towards negative returns and then begins to oscillate between positive 
and negative returns. These oscillations are clearly evident 
for an interval of $70$ trading days after the 1987 crash. In that 
case, this is the time interval the market needed to come 
back to a 'typical' state. This phenomenon is partly reflected 
into the oscillatory behavior of the 
Standard and Poor's 500 index (S\&P500) observed after the 1987 crash 
\cite{Sornette96}.

It is worth to investigate the changes of the ensemble return PDF 
not only by investigating the tails of the distribution but also 
its central part. In particular, it is important to understand 
whether in extreme days only the return mean value and the scale 
of the PDF are changed or if the shape of the PDF is modified also.   
To this end, we select the $9$ trading days of our database in which 
the S\&P500 has negative extreme returns. 
We also consider the opposite case of the $9$ trading days in which 
the S\&P500 has the greatest positive returns. 
These days are listed in Table I with the corresponding 
S\&P500 return value. In Figure 2 and 3 
we show the return distributions observed in the New York Stock 
Exchange in the days of extreme absolute returns. Specifically, 
Figure 2 shows the return distribution 
in crash days whereas Figure 3 shows the return distribution 
in rally days. Figure 2 shows that in crash days the PDF has 
a peak at a negative value of return. Moreover the distribution 
is asymmetric and the positive tail is steeper than the negative 
one. Therefore in crash days not only the scale but also 
the shape and symmetry properties of the distribution change. 
\begin{figure}[t]
\epsfxsize=2.6in
\epsfbox{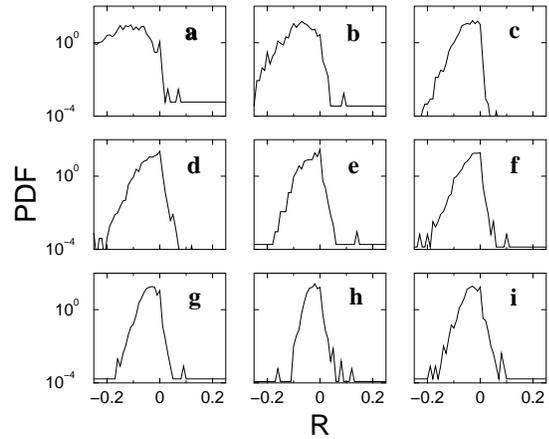}
\vspace{0.5cm}
\caption{Linear-log plot of the ensemble return distribution in days of 
S\&P500 index extreme negative return occurring in the investigated time
period (listed in the first part of Table I).}
\label{fig1}
 \end{figure}
  \begin{figure}[t]
\epsfxsize=2.6in
\epsfbox{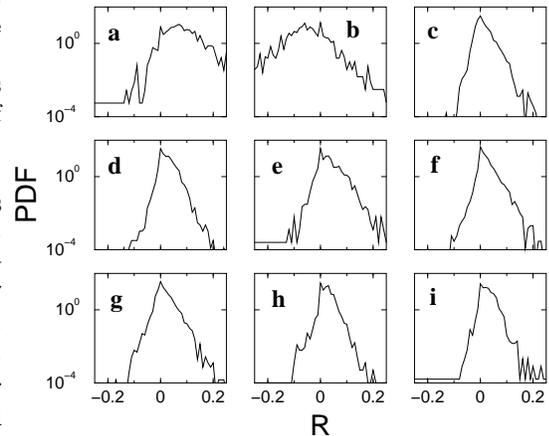}
\vspace{0.5cm}
\caption{Linear-log plot of the ensemble return distribution in days of 
greatest S\&P500 index positive return occurring in the investigated time
period (listed in the second part of Table I).}
\label{fig3}
 \end{figure}
A specular behavior is observed
during the days of great 
gain of the market. Figure 3 shows that in these trading days the 
negative tail of the distribution is steeper than the positive 
one and the distribution has a peak at a positive return. 
   
These findings can be quantified by noting that the distribution 
is negatively skewed in crash days, whereas the distribution is 
positively skewed in days of great gains. A quantitative estimate 
of the asymmetry of the PDF is difficult in finite statistical sets 
because the skewness parameter depends on the third moment of 
the distribution. Moments higher than the second are essentially 
affected by rare events rather than by the central part of the  
distribution. Due to the finite number of stocks in our statistical 
ensemble, a measure of the asymmetry of the distribution based 
on its skewness is not statistically robust. 
We overcome this problem by considering a different measure of 
the asymmetry of the distribution. Specifically, we extract the 
median and the mean of the distribution for all 
trading days. When a probability distribution 
is symmetric the median coincides with the mean. Therefore the
difference between the mean and the median is a measure
of the degree of asymmetry of the distribution. For 
positively (negatively) skewed distribution the median is smaller 
(greater) than the mean. The median depends weakly on the 
rare events of the random variable and therefore is much less 
affected than the skewness by the finiteness of the number 
of records of the ensemble. 
 In order to estimate the median value 
we construct an histogram of the returns and we evaluate the 
median value as the value for which the area of the histogram 
below and above it are equal.
       
Figure 4 shows the difference between the mean and the median as 
a function of the mean for each trading day of the investigated period.
In the Figure each circle refers to a different trading day.
The circles cluster in an asymmetrical pattern which resembles a
sigmoid shape. In days in which the mean is positive (negative) 
the difference between mean and median is positive (negative). 
In extreme days (for example those listed in Table I) the 
corresponding circles are characterized by great absolute value of 
the the mean and a great value of the difference between mean 
and median. 
Another result summarized in 
Figure 4 is that this effect is not exclusive of the days of extreme 
crash and rally but it is also evident for trading days 
of intermediate absolute average return.
The change of the shape and of the symmetry properties during 
the days of large absolute returns suggests that in extreme days 
the behavior of the market cannot be statistically described 
in the same way of the 'normal' periods. Moreover Figure 4 indicates
that the difference from normal to extreme behavior 
increases gradually with the absolute value of the average 
return.

Among the extreme days one circle does not cluster around the sigmoid
shape and shows a different behavior having a negative
mean but a positive difference between mean and median. This circle 
(indicated by an arrow in Figure 4) corresponds to the 20 October 
1987 day, which is the day after the black 
Monday. The ensemble return distribution for this day is shown 
in panel b of Figure 3. This day is quite anomalous 
because the S\&P500 had a $5.24\%$ positive return, but the 
mean return of all the assets traded in NYSE
was $-5.28\%$. In other words in this day companies performed returns
which were strongly correlated with their capitalization.   
In summary, with just one exception, our results provide an empirical 
evidence that the ensemble statistical properties of a set of stocks 
traded simultaneously in a financial market change in a systematic
way when the market moves far from the typical day characterized 
by a small average return.

We now compare the results of our empirical analysis with
the results obtained by modeling the stock price dynamics
with a simple model: the single-index model. The single-index 
model \cite{Elton,Campbell} assumes that the returns of all 
assets are controlled by one factor, usually called the market. 
 \begin{figure}[t]
\epsfxsize=2.6in
\epsfbox{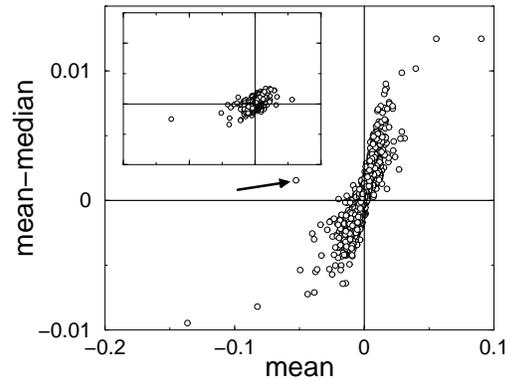}
\caption{Each circle is the difference between the mean and the median 
of the ensemble return distribution as a function of the mean for each 
trading day of the investigated time period.
The arrow indicates the anomalous behavior of the day after 
October 1987 black Monday.  
In the inset we show the same quantity for the artificial data generated 
according to the single-index model. The scale of the inset is the same
of the scale of the figure.}
\label{fig4}
 \end{figure}
For any asset $i$, we have
\begin{equation}
R_i(t)=\alpha_i+\beta_i R_M(t)+\epsilon_i(t),
\end{equation}

where $R_i(t)$ and $R_M(t)$ are the return of the asset $i$ and of the 
market at day $t$, respectively, $\alpha_i$ and $\beta_i$ are two 
real parameters and $\epsilon_i(t)$ is a zero mean noise term 
characterized by a variance equal to $\sigma^2_{\epsilon_i}$.  
The noise terms of different assets are uncorrelated, 
$<\epsilon_i(t) \epsilon_j(t)>=0$ for $i\neq j$. Moreover 
the covariance between $R_M(t)$ and $\epsilon_i(t)$ is set to
zero for any $i$.

Each asset is correlated with the market and the presence of such a 
correlation induces a correlation between any pair of assets. 
It is customary to adopt a broad-based stock index for the market $R_M(t)$. 
Our choice for the market is the Standard and Poor's 500 index.
The best estimation of the model parameters $\alpha_i$, $\beta_i$
and $\sigma^2_{\epsilon_i}$ is usually done with the ordinary least 
squares method \cite{Campbell}. To compare our empirical 
results with the predictions of the single-index model we 
build up an artificial stock market following Eq. (2). 
This is done by first evaluating the model parameters for all 
the assets traded in the NYSE and then by generating
a set of $n$ surrogate time series according to Eq. (2).
 
The ensemble return distribution computed in the artificial stock
market is symmetrical in 'typical' trading days.  In crash and rally days,
it is still approximately symmetrical around the mean value which 
can be positive (rallies) or negative (crashes). 
By contrast, as noted above, the return distribution in the real
ensemble is asymmetric in extreme days.
We can again quantify the asymmetry of the distribution by
evaluating the difference between mean and median. The inset of Figure 4 
shows the difference between the mean and the median as a function 
of the mean for the artificial data for each trading day of the 
investigated period. The differences between the real and artificial 
sets of circles are evident. The circles representing the synthetic 
data are distributed roughly symmetrically around the origin of the 
plane. Moreover the values of the difference between mean and median
observed for the single-index model are not very large compared with 
the ones observed in the real set, 
confirming that the ensemble return distribution of the 
artificial data is approximately symmetrical in extreme days too. 
This difference suggests that 
the effective correlation among the assets can be described 
by the single-index model only as a first approximation. The 
degree of approximation of the single-index model progressively
becomes worst for market days of increasing absolute average return
and fails in properly describing the market 
behavior of extreme days.

The main object of this letter is the study of the return distribution 
of an ensemble of stocks in a trading day with extreme absolute 
average return. 
We show that the ensemble return distribution 
changes shape and symmetry properties in crash and rally days.
We compare our empirical results with the expected behavior of the 
single-index model and we observe that this simple model fails 
in describing the market in extreme days. In particular the main 
discrepancy concerns the asymmetry of the ensemble return PDF  
of the model which is different from the one observed
in empirical data in extreme days.
Changes in the shape and symmetry of the PDF may be associated
to changes of the correlation properties.  
It is commonly accepted that the return time series of different 
stocks synchronously traded are correlated and several
researches has been performed in order to extract information from the 
correlation properties \cite{Mantegna99,Bouchaud99,Stanley99}. 
Our study suggests that the correlation properties between stocks 
may change during market days characterized by extreme absolute return.
A precise
characterization of the correlation properties and of their modification
is of key importance for the modeling of market dynamics in normal 
and in extreme market days.

The authors thank INFM and MURST for financial support. This work 
is part of the FRA-INFM project 'Volatility in financial markets'. 
F. Lillo acknowledges an FSE-INFM fellowships.
 We wish to thank Giovanni 
Bonanno for help in numerical calculations.



\begin{table}
\caption{List of the $18$ days of the investigated period (from January 
1987 to December 1998) in which the 
Standard and Poor's 500 index has the greatest return in absolute value.
The third column indicates the corresponding panel of the ensemble return 
distribution shown in Figures 2 and 3.}
\begin{tabular}{ccc}
Date&Standard and Poor's 500 return&Panel \\
\tableline
\tableline
19 10 1987&-0.2041&2a\\
26 10 1987&-0.0830&2b\\
27 10 1997&-0.0686&2c\\
31 08 1998&-0.0679&2d\\
08 01 1988&-0.0674&2e\\
13 10 1989&-0.0611&2f\\
16 10 1987&-0.0513&2g\\
14 04 1988&-0.0435&2h\\
30 11 1987&-0.0416&2i\\
\tableline
21 10 1987&+0.0908&3a\\
20 10 1987&+0.0524&3b\\
28 10 1997&+0.0511&3c\\
08 09 1998&+0.0509&3d\\
29 10 1987&+0.0493&3e\\
15 10 1998&+0.0418&3f\\
01 09 1998&+0.0383&3g\\
17 01 1991&+0.0373&3h\\
04 01 1988&+0.0360&3i\\
\end{tabular}
\end{table}

\end{document}